\documentclass[preprint,nofootinbib]{revtex4}%
\usepackage{amssymb}
\usepackage{amsfonts}
\usepackage{amsmath}
\usepackage{subfig}
\usepackage{graphicx}
\usepackage[usenames]{color}%
\providecommand{\U}[1]{\protect\rule{.1in}{.1in}}
\definecolor{blue}{rgb}{0,0,1}

\definecolor{red}{rgb}{1,0,0}

\begin{document}
\title{Instability of black strings in third-order Lovelock theory}
\author{Alex Giacomini$^{1}$, Carla Henr\'iquez-B\'aez$^{1}$, Marcela Lagos$^{2}$, Julio Oliva$^{2}$, Aldo Vera$^{2}$}
\affiliation{$^{1}$Instituto de Ciencias F\'{\i}sicas y Matem\'{a}ticas, Universidad
Austral de Chile, Casilla 567, Valdivia, Chile}
\affiliation{$^{2}$Departamento de F\'{\i}sica, Universidad de Concepci\'{o}n, Casilla
160-C, Concepci\'{o}n, Chile}
\begin{abstract}
We show that homogeneous black strings of third-order Lovelock theory are unstable under s-wave perturbations. This analysis is done in dimension $D=9$, which is the lowest dimension that allows the existence of homogeneous black strings in a theory that contains only the third-order Lovelock term in the Lagrangian. As is the case in general relativity, the instability is produced by long wavelength perturbations and it stands for the perturbative counterpart of a thermal instability. We also provide a comparative analysis of the instabilities of black strings at a fixed radius in general relativity, Gauss-Bonnet and third-order Lovelock theory. We show that the minimum critical wavelength that triggers the instability grows with the power of the curvature defined in the Lagrangian. The maximum exponential growth during the time of the perturbation is the largest in general relativity and it decreases with the number of curvatures involved in the Lagrangian.\footnote{alexgiacomini@uach.cl, carlah.baez@gmail.com, marcelagos@udec.cl, julioolivazapata@gmail.com, \\aldovera@udec.cl}

\end{abstract}
\maketitle

\tableofcontents

 \section{Introduction}

Gravity in dimensions greater than four has provided an interesting setup to
explore whether or not many of the properties that black holes have in four
dimensions are intrinsic to these kinds of spacetimes. According to string
theory, which provides a quantum theory containing a massless, spin-2 field,
the introduction of higher dimensions is actually a necessity and, in a
perturbative approach, Einstein-Hilbert action acquires corrections with
terms that contain higher powers in the curvature that modify the short distance dynamic of the
theory. Even more, Maldacena's conjecture \cite{Maldacena:1997re}, which was stated in
the realm of string theory and applies to the more general setup of
gauge/gravity duality, states that gravity in asymptotically anti de Sitter
spacetimes in $D$ dimensions is equivalent to a conformal field theory living in dimension $D-1$. These reasons have led in the last decades to a vast exploration of
gravity in higher dimensions and, in particular, to the study of black holes in
such scenarios (for a review, see \cite{horowitzbook}).

A family of configurations that are intrinsic to dimensions greater than four
are the asymptotically flat homogeneous black strings in general relativity.
Given the fact that the direct product of Ricci-flat manifolds is still Ricci-flat, one can add to the line element of the
Schwarzschild-Tangherlini spacetime in $d$ dimensions \cite{Tangherlini:1963bw}, the line
element of a locally flat space, obtaining black holes with horizon topologies
given by $S^{d-2}\times\mathbb{R}^{p}$ in $d+p$ dimensions, which receive the name of black strings in the case
$p=1$ and black $p$-branes in general. Without a cosmological constant, there
are no black holes in general relativity in three-dimensions ($d=3$);
therefore the black $p$-brane solutions exist only in dimensions greater than or equal to five, and they have to be added to the now vast zoo of solutions of
gravity in vacuum in higher dimensions that contains black rings \cite{Emparan:2001wn}, black
Saturns \cite{Elvang:2007rd}, and other configurations. Since these extended black
objects do Hawking radiate, it is natural to compare their entropy to
determine which of the phases of gravity in higher dimensions dominates for a
given mass. This comparison shows that there is a critical mass above which
the black strings have more entropy than the corresponding spherical black
hole and below which the black hole is the one possessing a larger entropy
\cite{Gregory:1993vy}. This led Gregory and Laflamme to look for the existence of a
perturbative instability of the black strings \cite{Gregory:1993vy}. Indeed, they showed that these configurations are linearly unstable against long wavelength
perturbations traveling along the black strings and $p$-branes. \footnote{These
unstable modes disappear if one compactifies the extended direction to a size
smaller than the minimum wavelength that triggers the instability, but it
reappears as soon as one reduces the mass of the black hole in the transverse
section.} Since this is a linearized perturbative instability, no dynamical
information can be extracted from the analysis to state which is the final
stage configuration. In order to form a set of black holes from the black
string the event horizon must pinch off, giving access to an external
observer to the singularity. Horowitz and Maeda \cite{Horowitz:2001cz} proved that such a
pinch off cannot occur in a finite affine parameter along the horizon
generator and therefore conjecture the existence of nonhomogeneous black
strings as possible end points of the Gregory-Laflamme (GL) instability. These
configurations were constructed perturbatively \cite{Gubser:2001ac} and numerically
\cite{Wiseman:2002zc}, but all of those happen to have less entropy than the homogeneous
black string, and therefore cannot be the end point of the instability. For
weakly nonhomogeneous solutions it was shown that there is a critical
dimension $D^{\ast}=13$ below which the homogeneous string carries more
entropy and above which the nonhomogeneous one has a larger entropy \cite{Sorkin:2004qq}
(later it was shown that by boosting the black string one can modify this
critical dimension \cite{Hovdebo:2006jy}). The question of the final stage of the GL
instability was open, until the series of works \cite{num0,num1}, which
culminate in the work by Lehner and Pretorius \cite{Lehner:2010pn}. After an
outstanding numerical effort, these works showed that the perturbed five dimensional black
string evolve towards a set of spherical black holes connected by black
strings and even more, for a finite time for an asymptotic observer, the radius
of the cylinders may shrink to zero and the system may than develop a null naked
singularity, providing a counterexample to cosmic censorship in five
dimensions without fine-tuning on the initial data. \footnote{See \cite{Figueras:2015hkb}
for a recent similar analysis and result in the case of the black ring.} No further study of the final point of the GL
instability was done in a different setup until recently, when the large $D$
expansion \cite{Emparan:2013moa} was used
to show that, at leading order in $1/D$, the black string instability leads to
a nonhomogeneous black string as a final stage \cite{Emparan:2015gva}, which is
consistent with the existence of a finite critical dimension $D^{\ast}$.

One expects that when the radius of the
horizon decreases enough, higher curvature terms might play a role in the
evolution of the perturbation and therefore it would be interesting to explore
the effect of higher curvature terms in the existence and evolution of black
strings. From all the possible higher curvature terms, there is a particular
family that is singled out by requiring second-order field equations as well
as diffeomorphism invariance. These terms are known as Lovelock terms \cite{Lovelock:1971yv} 
and are labeled by the order $n$ of the curvature. The cosmological
and the Einstein-Hilbert terms are the first two in the Lovelock family and
correspond to $n=0$ and $n=1$, respectively. The first nontrivial member of the Lovelock family is
quadratic in the curvature ($n=2$), is known as the Gauss-Bonnet term, and it
contributes to the field equations only in dimensions $D\geq5$. Something
similar occurs for the remaining terms of order $n$, which produce nontrivial
field equations only for dimension $D\geq2n+1$. Since these terms have second-order field equations, they are devoid of Ostrogradsky instabilities, and
therefore it is natural to explore their effect on the perturbative black
string instability even in regimes where they may completely dominate the dynamics. In the Gauss-Bonnet case different aspects of this problem
have been studied. Approximate black string solutions were constructed
numerically in \cite{Barcelo:2002wz}-\cite{Kobayashi:2004hq}. 
In particular in \cite{Brihaye:2010me} weakly nonhomogeneous static black strings were constructed for
Einstein-Gauss-Bonnet theory and the authors provided evidence for the
relation between perturbative and thermal instability in this scenario.\footnote{The existence of a possible critical dimension for the transition between these configurations and the homogeneous black strings, along the lines of the work by Sorkin \cite{Sorkin:2004qq}, is still an open problem.} For the
theory containing only the Gauss-Bonnet term, one can construct black strings
and $p$-branes by simply adding flat directions to the black holes found in \cite{Crisostomo:2000bb}. This was done simultaneously in Refs. \cite{KastorMann} and \cite{Giribet:2006ec} where it was shown as well that by
comparing entropies one must expect a GL instability for the black strings and
$p$-branes of sufficiently low ADM mass. Since perturbation theory in higher
curvature gravity presents subtleties (see, for example, \cite{Dotti:2005sq}-\cite{Gannouji:2013eka} for studies of
black hole perturbations), the perturbative search for the GL instability of
the solutions constructed in \cite{KastorMann}-\cite{Giribet:2006ec} has been only recently carried out in
\cite{Giacomini:2015dwa}.
There, it was shown that the black string in
Gauss-Bonnet theory (no Einstein term) in seven dimensions was unstable
under s-wave modes and that the instability is present above a certain
critical wavelength, in a similar fashion to general relativity.

Less is known for third-order Lovelock theory.\footnote{See \cite{R4} for a recent analysis on the effects on boosted black strings coming from fourth-order corrections in M-theory.} The spherically symmetric black
hole solution can be extracted from Wheeler's polynomial in \cite{Wheeler:1985qd}, which
turns out to be the unique spherically symmetric solution for generic values
of the coupling constants as shown by  Birkhoff's theorem \cite{Zegers:2005vx}. Some
early papers also explored the compactifications of the cubic Lovelock terms
\cite{MuellerHoissen:1985mm,MH2}, as well as some more recent papers constructing wormholes
\cite{Canfora:2008ka,Matulich:2011ct} or exploring the thermodynamics of
asymptotically AdS solutions in vacuum or with different types of matter
fields \cite{Hendi:2015psa} (see also \cite{Camanho:2013pda}
for a thorough analysis
of Wheeler's polynomial and some holographic considerations of third-order
Lovelock theory).

The existence of a perturbative black string instability for third-order
Lovelock gravity has not been explored at all, and it is
the purpose of this paper to partially contribute to that gap. As mentioned
above, in order to construct homogeneous black strings by simply adding a line
to a given black hole, one can consider the Lovelock terms separately. To
simplify the problem we consider the third-order Lovelock theory in nine
dimensions, which is the lowest dimension in which this can be done for $n=3$.
The analysis of \cite{KastorMann}-\cite{Giribet:2006ec} provided an entropic argument for the instability of
black strings with mass below a certain critical mass, and here we show that
such black strings are indeed perturbatively unstable against long wavelength
perturbations and that the dispersion relation for the instability is
qualitatively similar to the Gregory-Laflamme instability in general relativity. Even more, for a given
black hole radius in the transverse section of the black string, we compare
the maximum exponential growths and the critical wavelengths of the instabilities of
the solutions of general relativity, Gauss-Bonnet and third-order Lovelock
theory in nine dimensions. We find that the maximum critical wave number at which the instability is triggered decreases with the order of the Lovelock theory as well as the maximum exponential growth. This comparison is done for a fixed radius of
the horizon instead than for a fixed mass since the former is a geometric
quantity that can be defined independently of the dynamics that governs
gravity and is therefore suitable for comparing ``decay" processes in different theories.

Section II is devoted to the presentation of third-order Lovelock theory, its
black holes and the black strings, as well as some considerations on the thermodynamics.

In Sec. III we present the perturbative analysis for an s-wave on the
nine-dimensional black string for third-order Lovelock theory, and show that the string is unstable under long
wavelength perturbations. Using some scalar invariants, we show as well that
the unstable modes are physical and cannot be removed by gauge
transformations. In Sec. IV we provide a comparative analysis of the
dispersion relations for the instabilities of the black strings of Einstein,
Gauss-Bonnet and third-order Lovelock theory in nine dimensions. Section VI is
devoted to some comments and conclusions.

\bigskip

 \section{Lovelock theories}

 Lovelock gravity is the most general theory in higher dimensions (compatible with conservation of the stress-tensor) that leads to second-order field equations. The action of this theory is a sum of $[D/2]$ terms of the form
\begin{equation}
I_{(n)}= \frac{1}{2\kappa_D^2}\int d^Dx\sqrt{-g}\alpha_{n}\mathcal{L}_{(n)},
\end{equation}
where $\kappa_D$ and $\alpha_n$ are arbitrary constants which represent the coupling of the terms in the Lagrangian density given by
\begin{equation}
 \mathcal{L}_{(n)}=\frac{1}{2^n}\delta^{\mu_1\cdots\mu_n\nu_1\cdots\nu_n}_{\rho_1\cdots\rho_n
 \sigma_1\cdots\sigma_n}R_{\mu_1\nu_1}{}^{\rho_1\sigma_1}\cdots R_{\mu_n\nu_n}{}^{\rho_n\sigma_n}.
\end{equation}
From this, we can see that the first terms in $\mathcal{L}_{(n)}$ correspond to the
\begin{itemize}
 \item Cosmological term
 \begin{equation*}
  \mathcal{L}_{(0)}=1,
 \end{equation*}
 \item Einstein-Hilbert term
 \begin{equation*}
  \mathcal{L}_{(1)}=R,
 \end{equation*}
 \item Gauss-Bonnet term
  \begin{equation*}
   \mathcal{L}_{(2)}=R^2-4R_{\mu\nu}R^{\mu\nu}+R_{\mu\nu\rho\lambda}R^{\mu\nu\rho\lambda},
  \end{equation*}
  \item Third-order Lovelock term
  \begin{eqnarray*}
   \mathcal{L}_{(3)}=
   \begin{array}{c} 
   R^3-12RR_{\mu\nu}R^{\mu\nu}+16R_{\mu\nu}R^\mu{}_\rho R^{\nu\rho}+24R_{\mu\nu}R_{\rho\sigma}R^{\mu\rho\nu\sigma}+3RR_{\mu\nu\rho\sigma}R^{\mu\nu\rho\sigma}\\
   -24R_{\mu\nu}R^{\mu}{}_{\rho\sigma\kappa}R^{\nu\rho\sigma\kappa}+4R_{\mu\nu\rho\sigma}R^{\mu\nu\eta\xi}R^{\rho\sigma}{}_{\eta\xi}-8R_{\mu\rho\nu\sigma}R^{\mu}{}_\eta{}^{\nu}{}_{\xi} R^{\rho\eta\sigma\xi}.
  \end{array}
  \end{eqnarray*}
\end{itemize}
The field equations for an arbitrary linear combination of the terms $I_{(n)}$ are
\begin{equation}
\mathcal{E}_{\mu\nu}=\sum_{n=0}^{[D/2]}\alpha_{(n)}E_{\mu\nu}^{(n)}=0,
\end{equation}
where
\begin{equation}
 E^{\mu}{}_{\nu}^{(n)}=-\frac{1}{2^{n+1}}\delta^{\mu\eta_1\cdots\eta_n\xi_1\cdots\xi_n}_{\nu\rho_1\cdots\rho_n\sigma_1\cdots\sigma_n}R_{\eta_1\xi_1}{}^{\rho_1\sigma_1}\cdots R_{\eta_n\xi_n}{}^{\rho_n\sigma_n}.\label{eq:eom}
\end{equation}
Again, we have that the first terms in (\ref{eq:eom}) are
\begin{eqnarray*}
E^{(0)}_{\mu\nu}&=&-\frac{1}{2}g_{\mu\nu},\\
E^{(1)}_{\mu\nu}&=&R_{\mu\nu}-\frac{1}{2}Rg_{\mu\nu},\\
E^{(2)}_{\mu\nu}&=&2\left(RR_{\mu\nu}-2R_{\mu\rho}R^{\rho}{}_{\nu}-2R^{\rho\sigma}R_{\mu\rho\nu\sigma}+R_{\mu}{}^{\rho\sigma\gamma}R_{\nu\rho\sigma\gamma} \right)-\frac{1}{2}g_{\mu\nu}\mathcal{L}_{(2)},\\
E^{(3)}_{\mu\nu}&=&
 \begin{array}{c}
 3(R^2R_{\mu\nu}-4RR_{\rho\mu}R^{\rho}{}_{\nu}-4R^{\rho\sigma}R_{\rho\sigma}R_{\mu\nu}+8R^{\rho\sigma}R_{\rho\mu}R_{\sigma\nu}-4RR^{\rho\sigma}R_{\rho\mu\sigma\nu}\\
 +8R^{\rho\kappa}R^{\sigma}{}_{\kappa}R_{\rho\mu\sigma\nu}-16R^{\rho\sigma}R^{\kappa}{}_{(\mu}R_{|\kappa\sigma\rho|\nu)}+2RR^{\rho\sigma\kappa}{}_{\mu}R_{\rho\sigma\kappa\nu}+R_{\mu\nu}R^{\rho\sigma\kappa\eta}R_{\rho\sigma\kappa\eta}\\
 -8R^{\rho}{}_{(\mu}R^{\sigma\kappa\eta}{}_{|\rho|}R_{|\sigma\kappa\eta|\nu)}-4R^{\rho\sigma}R^{\kappa\eta}{}_{\rho\mu}R_{\kappa\eta\sigma\nu}+8R_{\rho\sigma}R^{\rho\kappa\sigma\eta}R_{\kappa\mu\eta\nu}-8R_{\rho\sigma}R^{\rho\kappa\eta}{}_{\mu}R^{\sigma}{}_{\kappa\eta\nu}\\
 +4R^{\rho\sigma\kappa\eta}R_{\rho\sigma\xi\mu}R_{\kappa\eta}{}^{\xi}{}_{\nu}-8R^{\rho\kappa\sigma\eta}R^{\xi}{}_{\rho\sigma\mu}R_{\xi\kappa\eta\nu}-4R^{\rho\sigma\kappa}{}_{\eta}R_{\rho\sigma\kappa\xi}R^{\eta}{}_{\mu}{}^{\xi}{}_{\nu})
 -\frac{1}{2}g_{\mu\nu}\mathcal{L}_{(3)}
\end{array}
\end{eqnarray*}
In what follows, we will consider theories with only one term in the action, namely, pure Lovelock theories, with $n\geq 1$, since in those theories it is possible to construct homogeneous black strings and black $p$-branes analytically \cite{Giribet:2006ec}.
\subsection{Black hole solutions in pure Lovelock theories}
According to \cite{Crisostomo:2000bb}, the spherically symmetric black hole solutions of Lovelock theories in $d$ dimensions with only the $n$th-order term in the action are of the form 
\begin{equation}
 ds_{\text{BH}}^2=-\left( 1-\left(\frac{r_{+}}{r} \right)^{\frac{d-2n-1}{n}}  \right)dt^2+\frac{dr^2}{1-\left(\frac{r_{+}}{r}  \right)^{\frac{d-2n-1}{n}}}+r^2d\Omega^2_{d-2}, \label{eq:BHS}
\end{equation}
where $d\Omega^2_{d-2}$ is the line element of a $(d-2)$-sphere and
$r_+$ corresponds to the radius of the event horizon which is related to the mass $m$ of the black hole by
\begin{equation}
 r_{+}^{d-2n-1}=\frac{2m\kappa_d^2(d-2n-1)!}{\alpha_n\Omega_{d-2}(d-2)(d-3)!}.
\end{equation}
\\ 
Below we will be interested in analyzing the evolution of perturbations on black strings that are obtained by oxidating these black holes. Given the fact that a $t=$const surface intersects the horizon at the bifurcation surface rather than at the future horizon it is useful to consider Kruskal-like coordinates, where the $T=$const surfaces do indeed intersect the future horizon, which will allow us to properly define the evolution of the perturbation.\\
Near to the horizon, the generalized tortoise coordinate $r^*$ for this family of black holes has the form
\begin{equation}
r^*\sim \frac{nr_+}{d-2n-1}\ln{(r-r_+)}.
\end{equation}
Then, we define the Kruskal null coordinates as
\begin{eqnarray*}
 U&=&-\exp\left(-\frac{d-2n-1}{2nr_+}(t-r^*)\right),\\
 V&=&\exp\left(\frac{d-2n-1}{2nr_+}(t+r^*)\right).
\end{eqnarray*}
And finally, we set
\begin{eqnarray*}
R=V-U,\qquad\qquad T=V+U.
\end{eqnarray*}
In terms of these coordinates, the metric (\ref{eq:BHS}) is regular at the horizon,  taking the form
\begin{equation}
 ds_{\text{BH}}^2\sim f^{\prime}(r_+)\left(\frac{nr_+}{d-2n-1}\right)^2\left(-dT^2+dR^2 \right)+r^2d\Omega^2_{d-2}
\end{equation}
and we call ($T,R$) the generalized Kruskal coordinates for this solution.\\
\subsection{Homogeneous black string solutions}
It is known that is possible to construct homogeneous black string and $p$-brane solutions in pure Lovelock theories with $D=d+p$, starting with the black hole solutions given by (\ref{eq:BHS}), making
\begin{align*}
ds^2=ds_{\text{BH}}^2+\sum_{i=1}^{p}dx_i^2,
\end{align*}
where the second term corresponds to a flat metric.
In particular, for $p=1$ we have the black string solution.
\bigskip

\subsection{Comparing entropies}
The entropies of the $D$-dimensional black hole and the (compactified) black string, in terms of the mass, have the following behavior:
\begin{equation}
 S_{\text{BH}}\sim m^{\frac{D-2n}{D-2n-1}},
\end{equation}
\begin{equation}
 S_{\text{BS}}\sim m^{\frac{D-2n-1}{D-2n-2}}.
\end{equation}
Then, given a theory we see that these entropies cross at a given critical mass $m_c$. For masses below $m_c$, the black hole solution is thermally favored.\\
The existence of the transition between the thermally favored solutions must be attached to a perturbative instability, according to the Gubser-Mitra conjecture \cite{Gubser:2000mm}.
\bigskip

\subsection{The s-wave perturbation}
As mentioned in the Introduction, the black strings in general relativity and Gauss-Bonnet theory, \cite{Gregory:1993vy} and \cite{Giacomini:2015dwa}, respectively, were proved to be unstable. In particular, the unstable mode in both cases corresponds to an s-wave mode, and therefore, inspired by this, we will consider the same family of perturbations in third-order Lovelock theory. Such a perturbation is defined by
\begin{equation}
H_{\mu\nu}\left(x^{\alpha}\right)=e^{\Omega t+ikz}h_{\mu\nu}(r),\label{eq:pert}
\end{equation}
where
\begin{equation*}
h_{\mu\nu}(r)=\begin{pmatrix}
h_{tt}(r) & h_{tr}(r) & 0 & 0\\
h_{tr}(r) & h_{rr}(r) & 0 & 0\\
0 & 0 & h(r)\sigma\left(S^{d-2}\right) & 0\\
0 & 0 & 0 & 0\\
\end{pmatrix}.
\end{equation*}

\bigskip
In the next section we will show that the black strings for third-order
Lovelock theory in nine dimensions are unstable under (\ref{eq:pert}).

\bigskip

\section{Black strings in third-order Lovelock theories are unstable}
We consider a homogeneous black string in $D=9$, the minimum dimension in which it is possible to construct this kind of solution in third-order Lovelock theory, and it is obtained from the oxidation of a black hole in $d=8$ where the gravitational potential is $f(r)=1-\left(\frac{r_+}{r}\right)^{1/3}$.\\
For convenience, we use a new radial coordinate $p=1-\left(\frac{r_+}{r}\right)^{1/3}$ which maps the region outside the horizon $r\in[r_+,+\infty[$ to $p\in[0,1[$.\\
In terms of $p$ the metric looks like
\begin{equation}
\label{eq:bsp}
ds^2=-pdt^2+\frac{9r_+^6}{p(1-p)^8}dp^2+\frac{r_+^6}{(1-p)^6}d\Omega^2_{(6)}+dz^2.
\end{equation}
\subsection{Linearized field equations}
From the linearized field equations of the perturbed metric we obtain the following expressions for the components $h_{tt}$, $h_{pp}$ and $h$ of $h_{\mu\nu}$ only in terms of $h_{tp}$,
\begin{align*}
h_{tt}(p)&=\frac{1}{9}\frac{p^2(p-1)^8}{\Omega r_+^2}\frac{d h_{tp}}{dp}+\frac{1}{9}\frac{p(p-1)^8}{\Omega r_+^2}h_{tp},\\
h_{pp}(p)&=-\frac{p}{\Omega}\frac{d^2h_{tp}}{dp^2}-2\frac{5p-1}{\Omega(p-1)}\frac{dh_{tp}}{dp}\\
         &\qquad+\frac{(-8p^8+56p^7-168p^6+280p^5-280p^4+168p^3-56p^2
+2[3k^2r_+^2+4]p+9\Omega^2r_+^2)}{p(p-1)^8\Omega}h_{tp},\\
h(p)&=\frac{p^2(p-1)^2}{12\Omega}\frac{d^2 h_{tp}}{dp^2}+\frac{p(p-1)(11p-3)}{12\Omega}\frac{dh_{tp}}{dp}+\frac{1}{12(p-1)^6\Omega}(9p^8-64p^7+196p^6\\
&\qquad-336p^5+350p^4-224p^3+84p^2-2[3k^2r_+^2+8]p-
9\Omega^2r_+^2+1)h_{tp},
\end{align*}
and where the $h_{tp}(p)$ component satisfies the following second-order master equation,
\begin{equation}\label{eq:mastereq}
A(p)\frac{d^2 h_{tp}}{dp^2}(p)+B(p)\frac{d h_{tp}}{dp}(p)+C(p)h_{tp}(p)=0,
\end{equation}
with
\begin{align*}
A(p)&=p^2(p-1)^8(p^8-8p^7+28p^6-56p^5+70p^4-56p^3+28p^2-8[3k^2r_+^2+1]p-
36\Omega^2r_+^2+1),\\
B(p)&=3p(p-1)^7(p^9+36\Omega^2r_+^2-9p^8+36p^7-84p^6+126p^5-126p^4+84p^3
-4[20k^2r_+^2+9]p^2\\
    &\qquad\qquad\qquad-[132\Omega^2r_+^2-16k^2r_+^2-9]p-1),\\
C(p)&=p^{16}-16p^{15}+120p^{14}-560p^{13}+1820p^{12}-4368p^{11}+8008p^{10}-2[57k^2r_+^2+5720]p^9\\
&\qquad+9[88k^2r_+^2-53\Omega^2r_+^2+1430]p^8+8[423\Omega^2r_+^2-294k^2r_+^2-1430]p^7\\
&\qquad+28[138k^2r_+^2-369\Omega^2r_+^2+286]p^6+84[210\Omega^2r_+^2-45k^2r_+^2-52]p^5\\
&\qquad+14[156k^2r_+^2-1305\Omega^2r_+^2+130]p^4+56[207\Omega^2r_+^2-12k^2r_+^2-10]p^3\\
&\qquad+12[6k^2r_+^2-357\Omega^2r_+^2+12k^4r_+^4+10]p^2+2[216\Omega^2k^2r_+^4+396\Omega^2r_+^2+3k^2r_+^2-8]p\\
&\qquad+324\Omega^4r_+^4-45\Omega^2r_+^2+1.
\end{align*}
Because an analytical solution of the master equation cannot be obtained, it is necessary first to analyse the asymptotic behavior of its solutions. Since Eq. (\ref{eq:mastereq}) is a second-order equation, 
there are two possible asymptotic behaviors near the horizon ($p=0$ or $r=r_+$), and with hindsight, the one that leads to a normalizable mode is given by
\begin{subequations}
\begin{eqnarray}
h_{tp}&=&p^{3m^3\Omega-1}(1+\mathcal{O}(p))\sim (r-r_+)^{3m^2\Omega-1},\label{asymprmas}\\
h_{tt}&=&\frac{1}{3m^3}p^{3m^3\Omega}(1+\mathcal{O}(p))\sim (r-r_+)^{3m^2\Omega},\\
h_{pp}&=&3m^3p^{3m^3\Omega-2}(1+\mathcal{O}(p))\sim (r-r_+)^{3m^2\Omega-2},\\
h&=&-\frac{m^3}{2\Omega}(k^2m^3+4\Omega+12\Omega^2m^3)p^{3m^3\Omega}(1+\mathcal{O}(p))\sim (r-r_+)^{3m^2\Omega}.
\end{eqnarray}
\end{subequations}
As it occurs in general relativity \cite{Gregory:1993vy} and Gauss-Bonnet theory \cite{{Giacomini:2015dwa}}, the two possible behaviors at infinity are given by an exponential function of the radial coordinate. 
For a given wavelength of the perturbation along the extended direction, $\lambda=k^{-1}$, we have to look for the values of $\Omega$ that allow us to smoothly connect the asymptotic behavior (\ref{asymprmas}) with an exponential 
decay at infinity. If it is possible to do so for some positive values of $\Omega$, we will say that the black string is unstable. Figure 1 shows that this is indeed the case. The details of the method and convergence considerations used to obtain these results are given in Appendix A.

\begin{figure}[h]
\begin{center}
\includegraphics[scale=0.3]{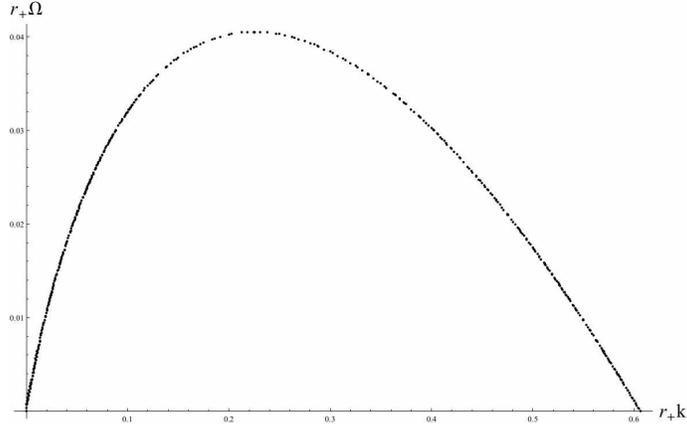}
\end{center}
\caption{Unstable modes of the black string in third-order Lovelock theory in $D=9$.}
\end{figure}

\subsection{Regularity conditions and gauge considerations}
From Fig. 1, it can be seen that the allowed values of $\Omega$ are such that $h_{tp}$ diverges at the horizon (see Eq. (\ref{asymprmas})). This happens as well for black strings in general relativity, and it led the 
authors of \cite{Gregory:1987nb} to state that there are no normalizable unstable modes in the case of an s-wave perturbation. Nevertheless, as explained in detail in \cite{Gregory:1994bj} (see also Chapter 2 
of \cite{horowitzbook}), given the fact that Schwarzschild coordinates do not cover the future horizon, it is necessary to write the perturbation in Kruskal coordinates, ($T,R$). Using the equations provided in Sec. II A, 
one can show that
\begin{equation*}
h_{TR}\sim(T-R)^{3m^3\Omega}.
\end{equation*}

Since the future horizon is located at the surface $R=T$ with $T>0$, we have proved that the unstable modes (with $\Omega>0$) are indeed finite on the domain of outer communications including the future horizon. The same occurs for the components $h_{RR}(T,R)$, $h_{TT}(T,R)$ and $h(T,R)$. 

It is interesting to see that the previous analysis has been done without imposing any gauge-fixing condition. Therefore, it is necessary to analyze whether or not the unstable modes we have found are physical. As was done for black strings in Gauss-Bonnet in seven dimensions \cite{{Giacomini:2015dwa}}, this can be done easily by considering the following scalar invariant:
\begin{equation}\label{eq:inv}
\mathcal{I}=709R_{\mu\nu\lambda\rho}R^{\lambda\rho}{}{}_{\sigma\tau}R^{\sigma\tau\mu\nu}
+890R^{\mu\nu}{}{}_{\lambda\rho}R^{\lambda\sigma}{}{}_{\nu\tau}R^{\rho\tau}{}{}
_{\mu\sigma}.
\end{equation}
This scalar identically vanishes on the unperturbed black string (\ref{eq:bsp}). Nevertheless (\ref{eq:inv}) is on-shell nonvanishing, on the perturbed black string\footnote{See \cite{Gibbons:2009dx} for a similar argument on nondiffeomorphic Einstein metrics on groups manifolds and coset spaces.} for $k>0$.
\\
Therefore we have proved that black strings in third-order Lovelock theory are unstable under long wavelength perturbations.

 \section{Comparing the black string instabilities in different theories}
We have proved that for the first three homogeneous terms on Lovelock theories, it is possible to construct homogeneous black strings that are unstable under gravitational perturbations on each theory. 
Given the wavelength of the perturbation as well as the rate of the exponential growth of the corresponding instability, for a given radius of the black hole in the transverse section of the string, one could 
wonder if it is possible to determine which is the dynamics that drove the instability. Such comparison makes sense only if kinematical quantities are involved, since the strength of the gravitational interaction 
on each theory depends on the value of the gravitational coupling, which might be different in each case. The radius of the horizon, being a purely geometrical quantity, provides us with a good parameter to perform 
the mentioned comparison. Now we present some details of the computations that allow one to determine which gravity theory destabilizes a black string of radius $r_+=1$ faster. 
\\
In order to fix the ideas, let us again consider a black string in $D=9$, for $n=1, 2$ and $3$, i.e., for pure Einstein, Gauss-Bonnet and third-order Lovelock theory, respectively. As mentioned above, the line elements 
for the unperturbed black strings can be obtained by oxidating the black holes given in Eq. (\ref{eq:BHS}). If we focus on the s-wave perturbations defined by (\ref{eq:pert}), one can find a master equation that, 
of course, takes the following form,
\begin{equation}\label{todas}
A_{n}\left(  r\right)  \frac{d^{2}h^{(n)}_{tr}}{dr^{2}}+B_{n}\left(  r\right)
\frac{dh^{(n)}_{tr}}{dr}+C_{n}\left(  r\right)  h^{(n)}_{tr}=0\ ,
\end{equation}
for $n=1,2,3$. The explicit expressions for each of the functions are given in Appendix B. Here $r$ is the Schwarzschild areal coordinate. Equation (\ref{todas}), at infinity, admits two possible asymptotic behaviors, 
one of which might provide a normalizable mode, since it decays exponentially. Near the horizon the asymptotic behaviors allowed by Eq. (\ref{todas}) are given 
by\footnote{Note that the factors that multiply $\Omega r_{+}$ in each case correspond to the inverse power of the decay in the gravitational potential of the black hole solution in the transverse section of the black strings.} 
\begin{equation}
 h^{(1)}_{tr}\sim(r-r_+)^{-1\pm\frac{1}{5}\Omega r_+},\qquad h^{(2)}_{tr}\sim(r-r_+)^{-1\pm\frac{2}{3}\Omega r_+},
 \qquad  h^{(3)}_{tr}\sim(r-r_+)^{-1\pm 3\Omega r_+} \ . 
\end{equation}
In an analogous manner to what we have discussed in the previous section, it can be shown that the ``plus" branch on each case provides a normalizable unstable ($\Omega>0$) physical mode, provided one deals with the 
perturbation using the Kruskal coordinates presented in Sec. II A. Figure 2 shows the unstable modes for a fixed radius simultaneously on the three different Lovelock theories that allow homogeneous black strings 
in $D=9$.
\bigskip

\begin{figure}[h]
\begin{center}
\includegraphics[scale=0.5]{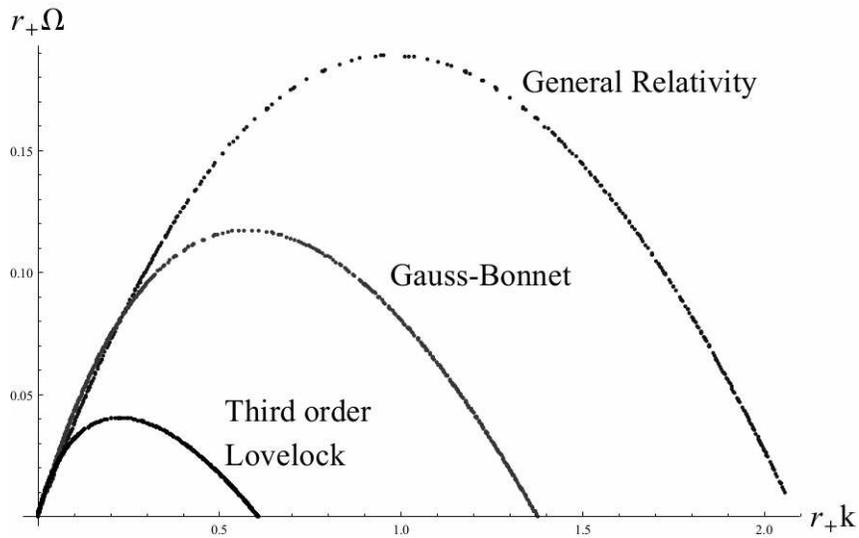}
\end{center}
\caption{Comparing unstable black strings for a given radius on Einstein, Gauss-Bonnet and third-order Lovelock theory.} 
\label{aba:fig1}
\end{figure}

From the picture we can see that the minimum critical wavelength that triggers the Gregory-Laflamme instability $\lambda_{c}^{(n)}$ grows with the power of the curvature involved on each theory. Therefore, for a fixed 
radius of the black hole in the transverse section of the string, if we compactify the $z$ direction to a scale $R_0\lesssim \lambda_{c}^{(1)}$, no unstable s-wave mode will be allowed. Then, if we start increasing $R_0$,
the black strings of general relativity become unstable first. Then the ones in Gauss-Bonnet theory become unstable for $R_0\gtrsim\lambda^{(2)}_{c}$, and finally the black strings of third-order Lovelock theory 
become unstable for $R_0\gtrsim\lambda^{(3)}_{c}$. The maximum exponential growth of the instability is obtained in general relativity, and it decreases with the order of the Lovelock theory.

\section{Conclusions}
We have shown that the homogeneous black strings of third-order Lovelock theory in $D=9$ are unstable for long wavelength perturbations traveling along the string. We have also provided a comparative analysis of the instabilities of homogeneous black strings in general relativity, Gauss-Bonnet and third-order Lovelock theory. By performing such comparison we have also proved that Gauss-Bonnet black strings in $D=9$ are unstable, extending our previous result obtained in $D=7$ in \cite{Giacomini:2015dwa}. It would be interesting to explore, in an approximate or numerical solution for an arbitrary linear combination of the three terms, how the dynamics of the perturbations of black strings are affected 
at different scales. These scales will be defined by the dimensionful coupling constants $\alpha_3$, $\alpha_2$ and $\alpha_1$. This would be important as well on the nonlinear evolution of the perturbations, since, 
as shown in \cite{Lehner:2010pn} in GR in $D=5$, the nonlinear evolution of a perturbation on a black string leads to a self-similar structure of black holes connected by thinner black strings, which ends on the formation of a 
null naked singularity. It is expected that higher curvature terms might play a role before the formation of such a singularity, which must occur before any effect of the backreaction of Hawking radiation plays some role.

\textbf{Acknowledgments}

The authors are grateful to Fabrizio Canfora and Gaston Giribet for enlightening comments.
A.V. appreciates the support of CONICYT Fellowship 21151067. 
M.L. appreciates the support of CONICYT Fellowship 21141229.
This work was supported by FONDECYT Grants No. 1141073 and No.
1150246. This project was also partially funded by Proyectos CONICYT, Research Council UK (RCUK) Grant No.
DPI20140053.

\section{Appendix}

\subsection{Numerical method and convergence}
To obtain the modes that fulfil the master equations for the perturbations we have used a power series solution. The method consists of imposing a solution that is a power series truncated to a given order $N$ around the horizon (which is a singular point of the master equation). Increasing the order $N$ one should obtain a better approximation of the full solution. When $N\rightarrow\infty$ the power series will certainly converge to the full solution at a given point $r=r_0$ if there are no other singular points on the complex plane in the ball centered at $r=r_+$ and with radius $|r_0-r_+|$. It is useful to work with coordinates $x$ that map the range $r\in[r_+,\infty[$ to $x\in[0,1[$. Then we proceed as follows: Fix a value for the momentum of the perturbation, $k=k_1$. Compute the power series solution around the horizon ($x=0$) to a given order $N$. Evaluate the power series solution at $x=1$, set it to zero, and find the real solutions of $\Omega$. Repeat these steps with the same value of $k$ but now truncating the series at order $N+1$ and stop the procedure after at least four significant figures of the obtained value of $\Omega$ are stabilized. Figures 3 and 4 provide evidence for the convergence of the process for $k=0.1$ and $k=0.5$, respectively, in the third-order Lovelock theory.
\begin{figure}[h]
\begin{center}
\includegraphics[scale=.35]{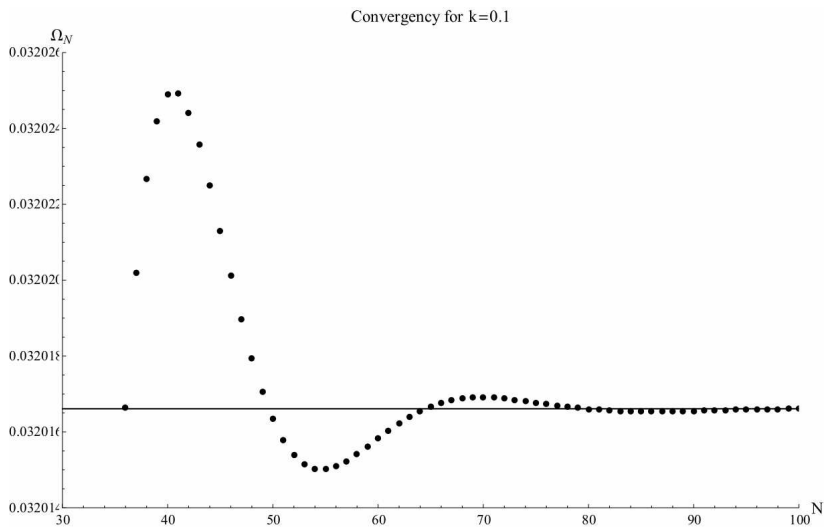}
\end{center}
\caption{Convergency of the frequencies obtained by the powers series solution for $k=0.1$.} 
\end{figure}
\begin{figure}[h]
\begin{center}
\includegraphics[scale=.35]{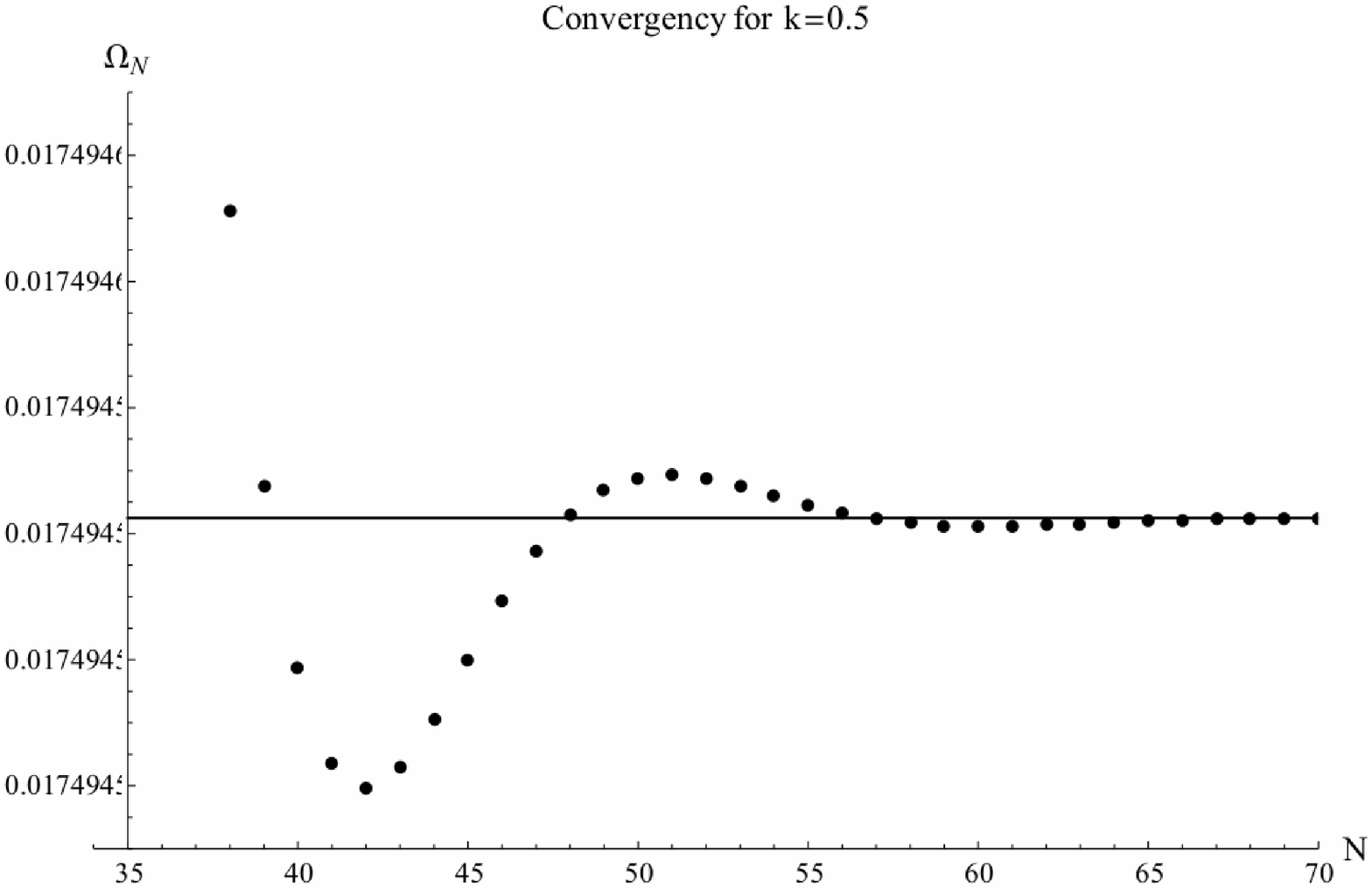} 
\end{center}
\caption{Convergency of the frequencies obtained by the powers series solution for $k=0.5$.} 
\end{figure}

It can be seen that convergency is achieved more slowly for smaller values of $k$.

\subsection{Master equation in different theories}

The explicit expressions for each of the functions in (\ref{todas}) are given by
\begin{itemize}
\item General relativity
\begin{eqnarray*}
A_1&=&r^2(r^5-r_+^5)^2(-4[\Omega^2+k^2]r^{12}+4k^2r_+^5r^7+25r_+^{10}),\\
B_1&=&r(r^5-r_+^5)(24[\Omega^2+k^2]r^{17}+4[9\Omega^2-2k^2]r_+^5r^{12}-16k^2r_+^{10}r^7-450r_+^{10}r^5+75r_+^{15}),\\
C_1&=&-(4[\Omega^2+k^2]^2r^{24}+24[\Omega^2+k^2]r^{22}-[8\Omega^2+k^2]k^2r_+^5r^{19}-4[12\Omega^2+23k^2]r_+^5r^{17}\\
&&+4k^2r_+^{10}r^{14}+[-101\Omega^2+137k^2]r_+^{10}r^{12}+1650r_+^{10}r^{10}-69k^2r_+^{15}r^7-1050r_+^{15}r^5+25r_+^{20} ).\\   
\end{eqnarray*}
\item Gauss-Bonnet theory
\begin{eqnarray*}
A_2&=&-16r(-2[5k^2+8\Omega^2]r^{11} -2[15k^2+8\Omega^2]r_+^3r^8  +9r_+^3r^6 +9r_+^6r^3\\
&&+2[16\Omega^2+15k^2]r_+^{\frac{3}{2}}r^{\frac{19}{2}} +10k^2r_+^{\frac{9}{2}}r^{\frac{13}{2}} -18r_+^{\frac{9}{2}}r^{\frac{9}{2}}  ),\\
B_2&=&8r(10[\Omega^2+5k^2]r^{10}-2[32\Omega^2-15k^2]r_+^3r^7-135r_+^3r^5-54r_+^6r^2 \\
&&-2[8\Omega^2+45k^2]r_+^{\frac{3}{2}}r^{\frac{17}{2}}+10k^2r_+^{\frac{9}{2}}r^{\frac{11}{2}}+189r_+^{\frac{9}{2}}r^{\frac{7}{2}}),\\
C_2&=&2r(-2[8\Omega^2+5k^2]^2r^{11} -40[5k^2+\Omega^2]r^9 -50k^4r_+^3r^8 \\
&&+5[8\Omega^2-177k^2]r_+^3r^6 -720r_+^3r^4 -72r_+^6r +20[8\Omega^2+5k^2]k^2r_+^{\frac{3}{2}}r^{\frac{19}{2}}\\
&&+80[9k^2+8\Omega^2]r_+^{\frac{3}{2}}r^{\frac{15}{2}} +365k^2r_+^{\frac{9}{2}}r^{\frac{9}{2}} +630r_+^{\frac{9}{2}}r^{\frac{5}{2}}).
\end{eqnarray*}
\item Third-order Lovelock theory
\begin{eqnarray*}
 A_3&=&   -972[\Omega^2+2k^2]r_+^{\frac{14}{3}}r^6 +27r_+^{\frac{14}{3}}r^4  -324[3\Omega^2+2k^2]r_+^4k^2r^{\frac{20}{3}} \\
 && +1944[\Omega^2+k^2]r_+^{\frac{13}{3}}r^{\frac{19}{3}} +648r_+^5k^2r^{\frac{17}{3}}  -54r_+^5r^{\frac{11}{3}} +27r_+^{\frac{16}{3}}r^{\frac{10}{3}} ,\\
 B_3&=&  -108[16k^2r_+^{\frac{14}{4}} +3\Omega^2r_+^{\frac{14}{3}}]r^5 +108r_+^{\frac{14}{3}}r^3 -432[3\Omega^2+2k^2]r_+^4k^2r^{\frac{17}{3}} \\
 && +540[3\Omega^2+4k^2]r_+^{\frac{13}{3}}r^{\frac{16}{3}} +432r_+^5k^2r^{\frac{14}{3}} -189r_+^5r^{\frac{8}{3}}  +81r_+^{\frac{16}{3}}r^{\frac{7}{3}} , \\
 C_3&=&  3( 144k^4r_+^{\frac{14}{3}}r^6 +3[338k^2+129\Omega^2]r_+^{\frac{14}{3}}r^4 +20r_+^{\frac{14}{3}}r^2 +36[3\Omega^2+2k^2]^2r_+^4r^{\frac{20}{3}} \\
 && -144[2k^2+3\Omega^2]k^2r_+^{\frac{13}{3}}r^{\frac{19}{3}} +144[3\Omega^2+2k^2]r_+^4r^{\frac{14}{3}} -72[12\Omega^2+13k^2]r_+^{\frac{13}{3}}r^{\frac{13}{3}}  \\
 && -366r_+^5k^2r^{\frac{11}{3}}  -28r_+^5r^{\frac{5}{3}} +9r_+^{\frac{16}{3}}r^{\frac{4}{3}}     )  .
\end{eqnarray*}
\end{itemize}

\end{document}